\documentclass[twocolumn,twoside]{revtex4}
\usepackage{graphicx}
\usepackage{fancyhdr}
\def\beq{\begin{equation}}
\def\eeq{\end{equation}}
\def\bea{\begin{eqnarray}}
\def\eea{\end{eqnarray}}
\def\nn{\nonumber}
\def\roughly#1{\mathrel{\raise.3ex\hbox
{$#1$\kern-.75em\lower1ex\hbox{$\sim$}}}}

\def\sla#1{\raise.15ex\hbox{$/$}\kern-.57em #1}

\def\bd{B^0_d}
\def\bdbar{{\bar B}^0_d}
\def\bs{B_s^0}
\def\bsbar{{\bar B}_s^0}
\def\btopik{B \to \pi K}

\def \mA0T{\mathcal{A}_{0, T}}

\begin{document}

\title{\boldmath CP Violation in $B$ Decays \\ (and the Search for New Physics)}

%

\author{David London}
\affiliation{Physique des Particules, Universit\'e de Montr\'eal, \\
C.P. 6128, succ.\ centre-ville, Montr\'eal, QC, Canada H3C 3J7}

\begin{abstract}

In this talk, I review the various measurements that have been made
over the years of CP-violating observables in $B$ decays. These
include indirect CP asymmetries, direct CP asymmetries, and triple
products. All are discussed in the context of the search for new
physics (NP). The only hints of NP have appeared in (i) $\btopik$
decays and (ii) 3-body $B \to K\pi\pi$ and $B \to K {\bar K}K$ decays,
where the extracted (loop-level) value of $\gamma$ may differ from the
value found using (tree-level) $B^+ \to D^{(*)} K^{(*)+}$ decays.
CP-violating observables may also be used to distinguish among the
models proposed to explain the current anomalies observed in $b \to s
\mu^+ \mu^-$ and $b \to c \tau {\bar\nu}_\tau$ decays.
\end{abstract}

\maketitle

\thispagestyle{fancy}


\section{Introduction}

The search for, and measurement of, CP violation in the $B$ system has
a long history ($\sim 30$ years). The goal was always to test the
Standard Model and (hopefully) find new physics (NP). In this talk, I
will review the subject of CP violation in $B$ decays, particularly
from the point of view of searching for NP.

CP violation always requires the interference of (at least) two
amplitudes with different weak (CP-odd) phases. CP violation in $B$
decays can be probed in three different ways:
\begin{enumerate}

\item Indirect CP violation. This involves $B_q^0$-${\bar B}_q^0$
  mixing ($q = d, s$). It arises due to the interference of the
  amplitudes $B^0_q \to f$ and $B^0_q \to {\bar B}^0_q \to f$. It
  requires a final state $f$ to which both $B_q^0$ and ${\bar B}_q^0$
  can decay. $f$ is usually (but not always) a CP eigenstate.

\item Direct CP violation. Here the signal is $\Gamma(B \to f) \ne
  \Gamma({\bar B} \to {\bar f})$. A nonzero direct CP asymmetry
  $A_{CP}$ requires that the interfering amplitudes have different
  weak and strong (CP-even) phases. Now, the strong phases are not
  known, which makes it difficult to use $A_{CP}$ to extract
  weak-phase information and test for NP. (There are some exceptions,
  which I'll describe.)

\item Triple products. Consider a decay $B \to f$, where $f$ involves
  (at least) 4 particles. In $|A(B \to f)|^2$, terms of the form
  ${\vec p}_1 \cdot ({\vec p}_2 \times {\vec p}_3)$, where the ${\vec
  p}_i$ are the 3-momenta of final-state particles, can appear. These
  are triple products (TPs). TPs are T violating, but not, in general,
  CP violating. A CP-violating effect can be found by comparing the
  TPs in $B \to f$ and ${\bar B} \to {\bar f}$. This CP-violating
  observable does not require that the interfering amplitudes have a
  strong-phase difference.

\end{enumerate}

\section{\boldmath $\alpha$, $\beta$, $\gamma$ and Tests of the SM}

\begin{figure}[!htbp]
\begin{center}
\includegraphics[width=0.4\textwidth]{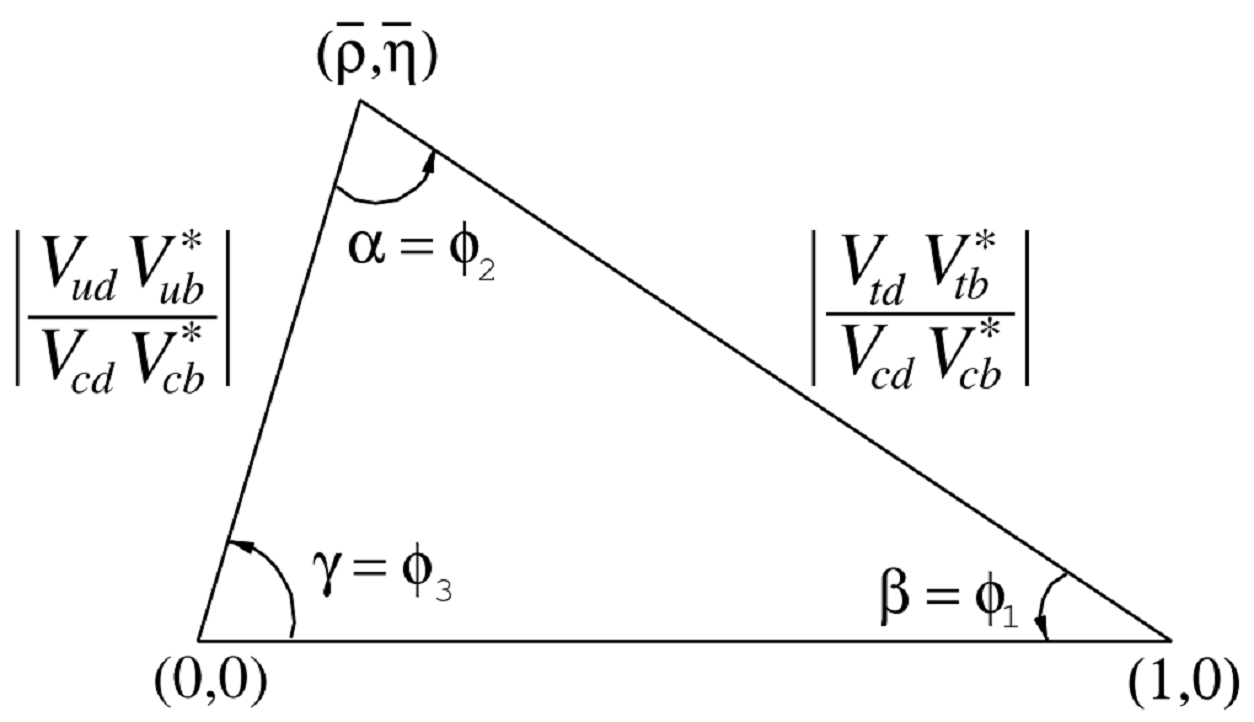}
\end{center}
\caption{\small The unitarity triangle.}
\label{UT}
\end{figure}

In the SM, the Cabibbo-Kobayashi-Maskawa (CKM) quark mixing matrix is
$3 \times 3$ and unitary. It is parametrized by three angles and one
phase. It is useful to write $V_{CKM}$ using the Wolfenstein
parametrization \cite{Wolfenstein:1983yz}:
\bea
V_{CKM} &=& 
\left(
\begin{array}{ccc} 
V_{ud} & V_{us} & V_{ub} \\ 
V_{cd} & V_{cs} & V_{cb} \\ 
V_{td} & V_{ts} & V_{tb} \\
\end{array}
\right) \\
&& \hskip-15truemm =~ 
\left(
\begin{array}{ccc} 
1 - \frac12 \lambda^2 & \lambda & A \lambda^3 (\rho - i \eta) \\ 
-\lambda & 1 - \frac12 \lambda^2 & A \lambda^2 \\ 
A \lambda^3 (1 - \rho - i \eta) & -A \lambda^2 & 1 \\
\end{array}
\right) + O(\lambda^4) ~. \nn
\eea
Writing $V_{ub} = |V_{ub}| e^{-i \gamma}$ and $V_{td} = |V_{td}| e^{-i
  \beta}$, the phase information of the CKM matrix can be encoded in
the unitarity triangle (Fig.~\ref{UT}), whose interior angles are
$\alpha$, $\beta$ and $\gamma$. The key point is that these angles can
be measured in $B$ decays; if it is found that $\alpha + \beta +
\gamma \ne \pi$, this is a sign of NP.

Originally, the idea was that these angles could be measured cleanly,
i.e., without any hadronic uncertainty, using indirect CP asymmetries
in the following decays:
\begin{itemize}

\item $\beta$: $\bd$ decays with ${\bar b} \to {\bar c}$, such as $\bd
  \to J/\psi K_S$.

\item $\alpha$: $\bd$ decays with ${\bar b} \to {\bar u}$, such as
  $\bd \to \pi^+\pi^-$.

\item $\gamma$: $\bs$ decays with ${\bar b} \to {\bar u}$, such as
  $\bs \to \rho^0 K_S$.

\end{itemize}
If these decays were dominated by their tree-level contributions, it
would be true that the CP phases could be extracted cleanly. However,
it was pointed out in Refs.~\cite{London:1989ph, Gronau:1989ia} that
all of these decays receive penguin contributions, which can spoil the
clean measurement. To be specific,
\begin{itemize}

\item $\bd \to J/\psi K_S$: to a good approximation, the ${\bar b} \to
  {\bar s}$ penguin diagram is real, as is the ${\bar b} \to {\bar c}
  c {\bar s}$ tree diagram.  Thus, there is effectively only a single
  amplitude with a weak phase, and $\beta$ can be extracted cleanly.

\item $\bd \to \pi^+\pi^-$: the weak phase of the ${\bar b} \to {\bar
  u} u {\bar d}$ tree diagram is $\gamma$, but the $t$-quark
  contribution to the ${\bar b} \to {\bar d}$ penguin diagram is
  $\beta$. The penguin contribution is subdominant, but it is not
  negligible. Thus, $\alpha$ can be extracted from $\bd \to
  \pi^+\pi^-$, but it is not a clean measurement: there is a
  theoretical uncertainty, of $O(|P/T|)$.

\item $\bs \to \rho^0 K_S$: as in $\bd \to \pi^+\pi^-$, the penguin
  diagram is subdominant, but not negligible. Similarly, $\gamma$ can
  be extracted from the indirect CP asymmetry in $\bs \to \rho^0 K_S$,
  but there is a theoretical uncertainty.

\end{itemize}

Regarding $\alpha$, it was shown in Ref.~\cite{Gronau:1990ka} that the
amplitudes for $\bd \to \pi^+\pi^-$, $B^+ \to \pi^+\pi^0$ and $\bd \to
\pi^0\pi^0$ are all related by isospin. By combining measurements of
all three decays, it is possible to remove the penguin pollution and
extract $\alpha$ cleanly.

As for $\gamma$, a new method was proposed \cite{Gronau:1990ra,
  Gronau:1991dp}. It involves the measurements of the rates of $B^+
\to D K^+$ and the CP-conjugate decays, where $D$ denotes $D^0$,
${\bar D}^0$ and $D_{CP}$ (which is identified by its decay to a
CP eigenstate such as $\pi^+\pi^-$). The rate difference between $B^+
\to D_{CP} K^+$ and $B^- \to D_{CP} K^-$ is a sign of direct
CP violation.  By combining the measurements of all the decays, the
strong phase can be determined and $\gamma$ can be extracted cleanly.
Variations of this method have been proposed \cite{Atwood:1996ci,
  Giri:2003ty}; all involve the decays $B^+ \to D^{(*)} K^{(*)+}$.

Using these methods, the results are \cite{CKMFitter}
\begin{itemize}

\item $\beta = \left( 22.14^{+0.69}_{-0.67} \right)^\circ$,

\item $\alpha = \left( 86.4^{+4.5}_{-4.3} \right)^\circ$,

\item $\gamma = \left( 72.1^{+5.4}_{-5.7} \right)^\circ$.

\end{itemize}
Combining the results gives
\beq
\alpha + \beta + \gamma = \left( 180.6^{+7.1}_{-7.2} \right)^\circ  ~.
\label{anglesresult}
\eeq
There is no sign of NP.

Note that, if one assumes that there are no tree-level NP
contributions to $B$ decays, then it is only NP in $\bd$-$\bdbar$
mixing that can affect the decays used for the extraction of $\alpha$,
$\beta$ and $\gamma$. That is, only $\bd \to \pi^+\pi^-$ and $\bd \to
J/\psi K_S$ can be affected. But the effects are in opposite
directions: if, due to the NP in the mixing, $\alpha_{\rm meas} =
\alpha_{\rm SM} + \phi_{\rm NP}$, then $\beta_{\rm meas} = \beta_{\rm
  SM} - \phi_{\rm NP}$, so that $\alpha_{\rm meas} + \beta_{\rm meas}
= \alpha_{\rm SM} + \beta_{\rm SM}$ \cite{Nir:1992wi}. That is,
$\alpha + \beta + \gamma = \pi$ was virtually assured.

In order to use $\alpha + \beta + \gamma \ne \pi$ as a test of NP, one
has to measure $\gamma$ using loop-level decays. NP will be indicated
if this loop-level value of $\gamma$ differs from the tree-level value
of $\gamma$ measured above. I will come back to this point.

So what do we learn from the result of Eq.~(\ref{anglesresult})? It
tells us nothing about NP in $\bd$-$\bdbar$ mixing -- that is
constrained by fits to the full CKM matrix \cite{CKMfits}. It does,
however, put constraints on tree-level NP. For example, it is clear
that large tree-level effects are ruled out. This is consistent with
the fact that no new particles have been ovserved at the LHC. On the
other hand, the error in Eq.~(\ref{anglesresult}) is $\sim 4\%$. While
this is small, it does leave room for NP, so it is worthwhile
continuing to explore CP-violating NP effects in $B$ decays.

Above, in the discussion of the penguin contribution to $\bd \to
J/\psi K_S$, I pointed out that, to a good approximation, the ${\bar
  b} \to {\bar s}$ penguin diagram is real. This means that the SM
also predicts that \cite{London:1997zk}
\beq
\beta ({\hbox{in}}~b \to c {\bar c} s) \simeq \beta ({\hbox{in ${\bar b} \to {\bar s}$ penguin decays}}) ~.
\eeq
This could be broken if there is NP in the ${\bar b} \to {\bar s}$
penguin. Historically, the experimental result is amusing. In 2003,
Belle extracted $\beta$ from $\bd \to \phi K_S$, finding
\cite{Abe:2003yt}
\beq
\sin 2\beta = -0.96 \pm 0.50 ^{+0.09}_{-0.11} ~,
\eeq
to be contrasted with its measurement of $\sin 2\beta = 0.73 \pm 0.06$
using $\bd \to J/\psi K_S$. This was a difference of 3.5$\sigma$, and
created a great deal of excitement. Unfortunately, with time the
effect went away.  The latest measurements give \cite{Amhis:2019ckw}
\bea
  \sin 2\beta ({\hbox{in}}~b \to c {\bar c} s) &=& 0.699 \pm 0.017 ~, \nn\\
 \sin  2\beta ({\hbox{in $b \to s$ penguin decays}}) &=& 0.648 \pm  0.038 ~.
\eea
The results agree, there is no sign of NP.

The SM predicts the weak phase of $\bs$-$\bsbar$ mixing to be
  very small, $O(10^{-2})$. The measured value is  \cite{Amhis:2019ckw}
\beq
\beta_s = 0.01848^{+0.00042}_{-0.00036} ~ {\hbox{radians}} ~,
\eeq
which is again consistent with the SM.

\newpage

\section{Direct CP Violation}

Consider the amplitude $A$ for the process $B \to f$. Suppose that $A = A_1 + A_2$, with
\beq
A_1 = |A_1| e^{i\phi_1} e^{i\delta_1} ~~,~~~~ A_2 = |A_2| e^{i\phi_2} e^{i\delta_2} ~,
\eeq
where the $\phi_i$ and $\delta_i$ are weak and strong phases,
respectively. The direct CP asymmetry is
\beq
A_{CP} \equiv \frac{|A|^2 - |{\bar A}|^2}{|A|^2 + |{\bar A}|^2} \propto \sin(\phi_1 - \phi_2) \sin(\delta_1 - \delta_2) ~.
\eeq
Thus, a nonzero $A_{CP}$ requires both nonzero weak-phase and
strong-phase differences.

This shows that, if a decay is dominated by a single weak amplitude in
the SM, $A_{CP}$ is predicted to vanish. This then provides a good way
to search for NP: the measurement of a nonzero direct CP asymmetry in
such a decay is a smoking-gun signal of NP. Unfortunately, there are
few decays in the SM that are dominated by a single amplitude. One is
$B^+ \to \phi K^+$, which is a pure ${\bar b} \to {\bar s}$ penguin
decay. Its direct CP asymmetry has been measured \cite{Amhis:2019ckw}:
$A_{CP} = 0.025 \pm 0.012$. So there is no clear sign of NP.

Conversely, if a decay has two non-negligible contributing diagrams,
with different weak phases, there may be a direct CP asymmetry.
Indeed, there are a great many measurements of 2-body $B$ decays, and
one finds some nonzero $A_{CP}$ results.  For example
\cite{Amhis:2019ckw}, $A_{CP}(B^+ \to \eta K^+) = -0.37 \pm 0.08$ and
$A_{CP}(B_s^0 \to \pi^+ K^-) = 0.213 \pm 0.017$. In the SM, these
decays receive contributions from two different amplitudes (tree and
penguin), with different weak phases, so that the appearance of a
nonzero $A_{CP}$ is not, in itself, surprising. But because the strong
phases cannot be calculated, we don't know what the SM prediction
is. Thus, in general, one cannot use measurements of direct CP
violation to test for NP.

There are exceptions. One is the use of $B^+ \to D^{(*)} K^{(*)+}$
decays to extract $\gamma$. Another is $B \to\pi K$ decays. Here there
are four decays: $B^+ \to \pi^+ K^0$, $B^+ \to \pi^0 K^+$, $\bd \to
\pi^- K^+$ and $\bd \to \pi^0 K^0$. The amplitudes obey a
quadrilateral isospin relation:
\beq
\sqrt{2} A^{00} + A^{-+} = \sqrt{2} A^{0+} + A^{+0} ~.
\eeq

Nine observables have been measured using these decays: 4 branching
ratios, 4 direct CP asymmetries $A_{CP}$, and 1 indirect CP asymmetry
$S_{CP}$ in $\bd\to \pi^0K^0$. These decays involve several
contributing amplitudes (diagrams) \cite{Gronau:1994rj,
  Gronau:1995hn}.  Keeping only the leading-order diagrams, one
expects 
\beq
A_{CP}(B^+ \to \pi^0 K^+) = A_{CP}(\bd \to \pi^- K^+) ~. 
\eeq
However, measurements give \cite{Amhis:2019ckw}
\bea
A_{CP}(B^+ \to \pi^0 K^+) &=& 0.040 \pm 0.021 ~, \nn\\
A_{CP}(\bd \to \pi^- K^+) &=&  -0.082 \pm 0.006 ~,
\eea
which differ by $5.5\sigma$. This is the ``$\btopik$ puzzle.''

The obvious question is then: is there really a disagreement with the
SM, or is this naive discrepancy misleading? Over the years, there
have been many analyses of the $\btopik$ puzzle, both within the SM
and with NP.

Recently, my collaborators and I reexamined this question
\cite{Beaudry:2017gtw}. We wrote the amplitudes in terms of five
diagrams: $T'$, $C'$, $P'_{tc}$, $P'_{EW}$ and $P^{\prime C}_{EW}$
(the small $P'_{uc}$ diagram was neglected). Within the SM, to a good
approximation, $P'_{EW}$ and $P^{\prime C}_{EW}$ can be related to
$T'$ and $C'$ using flavour SU(3) symmetry
\cite{EWPtree1,EWPtree2,EWPtree3}. Thus, there are three independent
diagrams, leading to 6 unknown parameters: the magnitudes and relative
strong phases of $T'$, $C'$ and $P'_{tc}$, and the weak phase
$\gamma$. With 9 observables, we can do a fit. (This is how we can
extract information despite unknown strong phases.)

There are two issues in the fit. First, theoretically, $C'$ is indeed
subdominant: the preferred value is $|C'/T'| = 0.2$. Second, as
discussed previously, there is an independent measurement of $\gamma$:
$\gamma = \left( 72.1^{+5.4}_{-5.7} \right)^\circ$.  This can be added
as input to the fit, or the value of $\gamma$ can be extracted. Note
that, if we extract $\gamma$ from $\btopik$ decays, this gives us a
loop-level value.

Here are the results:
\begin{enumerate}

\item If we fix $|C'/T'| = 0.2$ and set $\gamma$ to its measured
  value, we get a very poor fit (p value $=0.03$).

\item If we fix $|C'/T'| = 0.5$, which is theoretically on the large
  side and set $\gamma$ to its measured value, we get a good fit (p
  value $=0.43$).

\item If we fix $|C'/T'| = 0.5$ and allow $\gamma$ to vary, we still
  get a decent fit (p value $=0.36$). However, the best-fit
  (loop-level) value of $\gamma = (51.2 \pm 5.1)^\circ$ disagrees with
  the measured (tree-level) value by $2.7\sigma$.

\end{enumerate}
The conclusion is that the situation is still unclear. There is still
some unresolved tension in $\btopik$ decays. Hopefully, Belle II will
be able to shed some light on this.

The 3-body $B$ decays $B \to K \pi \pi$, $K {\bar K} K$, $\pi \pi \pi$
and $K {\bar K} \pi$ have also been studied. The direct CP asymmetries
have been measured, both at the inclusive level and in localized areas
of the Dalitz plot \cite{Aaij:2013sfa, Aaij:2013bla}, and nonzero
values of $A_{CP}$ have been found. However, because we have no
information about the strong phases, we run into the same problem as
we did with 2-body decays. Although these results have been analyzed
with particular theoretical models \cite{Xu:2013dta, Bediaga:2013ela,
  Cheng:2013dua, Lesniak:2013tsa, Wang:2014ira, Cheng:2016shb,
  Klein:2017xti}, and with U-spin symmetry \cite{Bhattacharya:2013cvn,
  Xu:2013rua}, there have been no unambiguous signals of NP.

But here too, there is an exception, similar to $\btopik$ decays. It
is possible to extract the loop-level value of $\gamma$ without
hadronic uncertainties from $B_d^0 \to K^+ \pi^0 \pi^-$, $B_d^0 \to
K_S \pi^+ \pi^-$, $B_d^0 \to K_S K_S K_S$, $B_d^0 \to K^+ K_S K^-$ and
$B^+ \to K^+ \pi^+ \pi^-$. The ingredients that go into this analysis
are as follows:
\begin{itemize}

\item In charmless $B \to PPP$ decays ($P$ is a pseudoscalar), the
  final states are related by flavour SU(3) symmetry. Because there
  are three (identical) final-state particles, there are six symmetry
  states \cite{Lorier:2010xf, Imbeault:2010xg}. The method uses the
  fully-symmetric (FS) state \cite{Bhattacharya:2014eca}.

\item In 3-body decays, one measures the Dalitz plots (DPs). For a
  particular process, it is possible to construct its decay amplitude
  from the DP by using the isobar method. From this, one can obtain the
  FS amplitude. For all $B \to PPP$ decays, the FS amplitudes are
  written in terms of (3-body) diagrams \cite{Lorier:2010xf}.

\item In Ref.~\cite{ReyLeLorier:2011ww}, the method for extracting
  $\gamma$ from $B_d^0 \to K^+ \pi^0 \pi^-$, $B_d^0 \to K_S \pi^+
  \pi^-$, $B_d^0 \to K_S K_S K_S$, $B_d^0 \to K^+ K_S K^-$ and $B^+
  \to K^+ \pi^+ \pi^-$ was proposed. Under SU(3), the amplitudes for
  these decays are all written in terms of the same four effective
  diagrams.  Thus, there are 9 theoretical parameters: the magnitudes
  and relative strong phases of these diagrams, an SU(3)-breaking
  parameter $\alpha_{\rm SU(3)}$, and the weak phase $\gamma$.  Given
  the FS amplitudes for these decays (from the isobar method), one can
  construct 11 FS observables (branching ratios, direct and indirect
  CP asymmetries) at each point of the DP. With more observables than
  theoretical parameters, a fit can be performed, permitting the
  extraction of $\gamma$.

\item Note that all parameters will vary throughout the DP, {\it
  except $\gamma$}. This holds also for $\alpha_{\rm SU(3)}$. We
  expect it to be positive at some points, and negative at
  others. Thus, by averaging over all points of the DP, the size of
  SU(3) breaking may be reduced.

\end{itemize}

Using the published measurements by BaBar of the DPs for these decays
\cite{Aubert:2008bj, Aubert:2009me, BABAR:2011ae, Lees:2011nf,
  Lees:2012kxa}, a preliminary implementation of the method was
carried out in Ref.~\cite{Bhattacharya:2013cla}. This analysis was
redone in Ref.~\cite{Bertholet:2018tmx}, taking into account all
systematic errors, particularly those from correlations between the
results from different DP points. It was found that (i) when averaged
over the full DP, SU(3) breaking is only at the percent level, and
(ii) the loop-level $\gamma$ can be extracted with a 6-fold ambiguity:
\begin{enumerate}

\item $\left( 12.9\mbox{}\,^{+8.4\phantom{1}}_{-4.3}   \text{ (stat)}  \pm 1.3 \text{ (syst)} \right)^\circ$, 

\item $\left( 36.6\mbox{}\,^{+6.6\phantom{1}}_{-6.1}  \text{ (stat)}  \pm  2.6 \text{ (syst)} \right)^\circ$,

\item $\left( 68.9\mbox{}\,^{+8.6\phantom{1}}_{-8.6}  \text{ (stat)}  \pm  2.4 \text{ (syst)} \right)^\circ$, 

\item $\left( 223.2\mbox{}\,^{+10.9}_{-7.5}  \text{ (stat)} \pm  1.0 \text{ (syst)} \right)^\circ$,

\item $\left(  266.4\mbox{}\,^{+9.2}_{-10.8}  \text{ (stat)} \pm  1.9 \text{ (syst)} \right)^\circ$, 

\item $\left(  307.5\mbox{}\,^{+6.9\phantom{1}}_{-8.1}  \text{ (stat)}  \pm  1.1 \text{ (syst)} \right)^\circ$.

\end{enumerate}
From these results, at this stage we do not know if the loop-level
value of $\gamma$ agrees with the tree-level value: \#3 is consistent,
while the others are not. This ambiguity can, in principle, be removed
by repeating the analysis with other symmetries of the final state
(mixed, antisymmetric). This is work in proogress.

\section{Triple Products}

Consider the decay $B \to f$, with $f$ decaying to four
particles. Suppose that there are two contributions to the decay
amplitude, $A = A_1 + A_2$, with $A_1 = |A_1| e^{i\phi_1}
e^{i\delta_1}$ and $A_2 = |A_2| e^{i\phi_2} e^{i\delta_2}$. $|A|^2$
may contain the term $\epsilon_{\mu\nu\rho\sigma} p_1^\mu p_2^\nu
p_3^\rho p_4^\sigma$, where the $p_i$ are the 4-momenta of the
final-state particles. Using energy-momentum conservation, this can be
written $\epsilon_{\mu\nu\rho\sigma} p_B^\mu p_2^\nu p_3^\rho
p_4^\sigma$, so that, in the rest frame of the $B$, this is $m_B \, {\vec
  p}_2 \cdot ({\vec p}_3 \times {\vec p}_4)$. This is called a triple
product (TP).

TPs are T-odd terms. The coefficient of the TP is
\bea
{\rm Im}(A_1 A_2^*) &=& |A_1||A_2| (
\underbrace{ \sin(\phi_1 - \phi_2) \cos(\delta_1 - \delta_2)}_{{\hbox{\scriptsize true TP}}} \nn\\
&& 
+~\underbrace{ \cos(\phi_1 - \phi_2) \sin(\delta_1 - \delta_2)}_{{\hbox{\scriptsize fake TP}}} ) ~.
\eea
The piece labeled `true TP' (`fake TP') is CP-odd (CP-even). From
this, we see that a TP is not, by itself, a signal of CP violation --
it can be nonzero even if $\phi_1 - \phi_2 = 0$ (as long as $\delta_1
- \delta_2 \ne 0$).

In order to obtain a CP-violating signal, one must compare the TP in
$B \to f$ with that in ${\bar B} \to {\bar f}$. Now, TPs are
themselves coefficients of P-odd terms in the angular distribution.
This implies that, given the TP in $B \to f$, one obtains the TP in
${\bar B} \to {\bar f}$ by changing the sign of the weak phases and
multiplying by $-1$ (due to the P-odd term). The upshot is that a true
CP-violating is obtained by {\it adding} the TPs in $B \to f$ and
${\bar B} \to {\bar f}$ \cite{Datta:2003mj, Gronau:2011cf}.  TPs are
sensitive to NP in the decay.

The most common type of decay used to look for TPs is $B \to V_1 (\to
P_1 P'_1) V_2 (\to P_2 P'_2)$, where the $V_i$ and $P_i^{(\prime)}$
are vector and pseudoscalar mesons, respectively. Examples include (i)
decays which are $b \to s {\bar s} s$ at the quark level: $\bd \to
\phi K^*$ (BaBar, Belle, LHCb) \cite{Amhis:2019ckw} and $\bs \to \phi
\phi$ (LHCb) \cite{Aaij:2019uld}, and (ii) decays which are $b \to s {\bar
  d} d$ at the quark level: $\bs \to K^*(892) {\bar K}^*(892)$ (LHCb)
\cite{Aaij:2017wgt}. In $B \to V_1 V_2$ decays, only the three vector
helicity amplitudes $A_0$, $A_\parallel$ and $A_\perp$ are
involved. Furthermore, because both $V_1$ and $V_2$ decay to two
pseudoscalars, there are only two TPs, proportional to ${\rm
  Im}[A_\perp A_0^*]$ and ${\rm Im}[A_\perp A_\parallel^*]$. (More
complicated decays, such as those involving fermions in the final
state, have three TP terms, as we will see below.) In experiments that
measure these TPs, it is important to allow for the possibility of NP
contributions and to take into account the background from other
(scalar) particles that also decay to $P_1 P'_1 P_2 P'_2$. This is
done in Ref.~\cite{Bhattacharya:2013sga}.

In all of the above decays, there is a single dominant decay amplitude
in the SM, so that the SM predicts no CP-violating TPs. And indeed,
this is what is found. For example, in $\bs \to \phi \phi$ (LHCb)
\cite{Aaij:2019uld}, the TPs whose coefficients are ${\rm Im}[A_\perp
  A_\parallel^*]$ and ${\rm Im}[A_\perp A_0^*]$ are called $A_U$ and
$A_V$, respectively.  It is found that
\beq
A_U = -0.003 \pm 0.015 ~~,~~~~ A_V = -0.012 \pm 0.015 ~.
\eeq
Thus, there is no evidence of NP in the decays (i) $b \to s {\bar s}
s$ or (ii) $b \to s {\bar d} d$.

\section{\boldmath $B$ Anomalies and TPs}

Another decay with four particles in the final state is $B \to K^*
(\to K \pi) \mu^+ \mu^-$. Of course, this is better known as one of
the pieces of the $b \to s \mu^+ \mu^-$ anomalies \cite{Isidori}. To
be specific, the angular distribution in $B \to K^* \mu^+ \mu^-$
contains $P'_5$, whose measurement disagrees with the prediction of
the SM.

Many NP models have been proposed to explain the $b \to s \mu^+ \mu^-$
anomalies. The question is: how can we distinguish among them? One way
is to use the fact that different NP models may predict different
CP-violating effects, namely TPs. The $B \to K^* \mu^+ \mu^-$ angular
distribution, which is $q^2$-dependent, involves 12 angular functions.
9 of them are CP-conserving (including $P'_5$), while 3 of them are TP
terms. Their (precise) measurement can help to differentiate NP
models.

\begin{figure}[!htbp]
\begin{center}
\includegraphics[width=0.7\hsize]{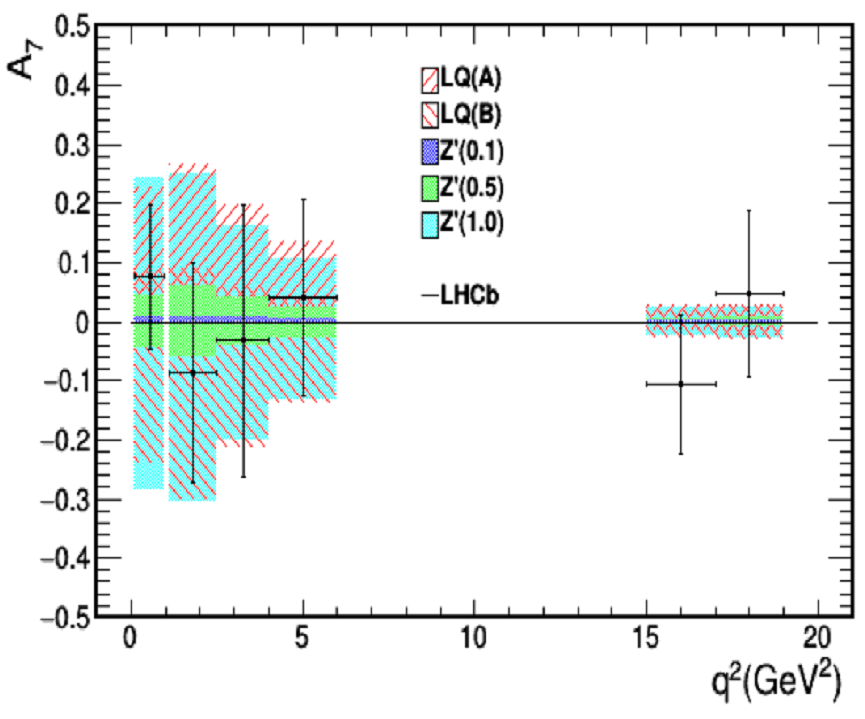}
\end{center}
\caption{\small Predictions of various LQ and $Z'$ models for the TP
  asymmetry $A_7$ in $B \to K^* \mu^+ \mu^-$.}
\label{bsmumuTP}
\end{figure}

This was studied in Ref.~\cite{Alok:2017jgr}. By comparing the $B \to
K^* \mu^+ \mu^-$ angular distribution with that of the antiprocess,
one can construct both direct CP asymmetries and TP asymmetries. Of
these, the TP asymmetries, $A_7$, $A_8$ and $A_9$, are the most
interesting, since they do not depend on the strong phases. And of
these, $A_7$ is the most promising, as it depends on the phase of the
Wilson coefficient $C_{10}$, which appears in fits to the data to
receive NP contributions. The NP models proposed to explain the $b \to
s \mu^+ \mu^-$ anomalies all involve either a leptoquark (LQ) or a
$Z'$ boson. In Ref.~\cite{Alok:2017jgr}, the contributions of LQ and
$Z'$ models to the TP asymmetries were computed for different ranges
of NP parameters. The results for $A_7$ are shown in
Fig.~\ref{bsmumuTP}. If this TP is found to be large, say 15-30\%,
this can only be explained within a subset of possible NP models.

There are also hints of NP in $b \to c \tau {\bar\nu}_\tau$
\cite{Sandilya}. Here too there are a variety of NP models that have
been proposed to explain the data. Once again, we would like to be
able to distinguish among them. Now, the decay $B \to D^* (\to D \pi)
\ell {\bar\nu}_\ell$ contains four final-state particles, so one
immediately thinks of TPs.

The case of $\ell = \mu$ is studied in
Ref.~\cite{Bhattacharya:2019olg}, using a model-independent
effective-field-theory approach. Here there are a total of 7 NP
parameters. It is found that, like $B \to K^* \mu^+ \mu^-$, the
angular distribution involves 12 angular functions. 9 CP-conserving
and 3 CP-violating (TP terms). Since the SM predicts no CP violation
in this decay, the measurement of a nonzero TP term would be a
smoking-gun signal of NP. Furthermore, it is possible to extract all 7
NP parameters, so that, if NP were found, we would obtain important
information about its properties.

Why examine $\ell = \mu$? There are two reasons. First, although the
NP is hinted at in $b \to c \tau {\bar\nu}_\tau$, that same NP might
affect $b \to c \mu {\bar\nu}_\mu$. Second, Belle has already measured
the angular distribution in $B \to D^* (\to D \pi) \mu {\bar\nu}_\mu$
\cite{Abdesselam:2017kjf}. The purpose was to extract $|V_{cb}|$, not
to look for NP, so TP terms were not included in their angular
distribution. With the results of Ref.~\cite{Bhattacharya:2019olg},
they can redo the analysis to search for NP.

Turning to $\ell = \tau$, unfortunately, it is not possible to measure
the angular distribution. Since the final-state particles in $\tau$
decay include an undetected neutrino, the 4-momentum of the $\tau$
cannot be measured. On the other hand, if one also considers the decay
$\tau^- \to \pi^- \nu_\tau$, so that the full decay is $B \to D^* (\to
D \pi') \tau^- (\to \pi^- \nu_\tau) {\bar\nu}$, one can construct an
angular distribution, including TP terms, using the final-state
particles $D$, $\pi'$ and $\pi^-$ \cite{Bhattacharya:2020lfm}.

In this case, the angular distribution depends on $q^2$ and $E_\pi$,
the energy of the $\pi^-$ in the $\tau$ decay. Even so, there are some
similarities with the NP analysis of $B \to D^* (\to D \pi) \mu
{\bar\nu}_\mu$: there are still 7 NP parameters, and the angular
distribution again contains 12 angular functions, 9 CP-conserving and
3 CP-violating (TP terms). Measuring this angular distribution is more
challenging, but if it can be done, the 7 NP parameters can be
extracted, providing an important first step towards the
identification of the NP.

\section{Conclusions}

In the 1990s, CP violation in the $B$ system was studied in great
detail, in anticipation of the measurements to come from BaBar and
Belle. The hope was that the measurement of indirect CP asymmetries
would reveal large NP effects. Unfortunately, that did not happen --
all measurements seem to be consistent with the SM.

Using CP-violating measurements, the only hints of NP have appeared
(surprisingly) in observables that involve direct CP violation. To be
specific, the value of $\gamma$ extracted from processes involving
loop-level penguin contributions -- (i) $\btopik$ decays and (ii)
3-body $B \to K\pi\pi$ and $B \to K {\bar K}K$ decays -- may be
different from the value found using tree-level $B^+ \to D^{(*)}
K^{(*)+}$ decays.

At present, there are (CP-conserving) hints of NP in processes
involving $b \to s \mu^+ \mu^-$ and $b \to c \tau {\bar\nu}_\tau$
decays. In both cases, several NP models have been proposed to explain
the data.  To differentiate among the various models, measurements
of CP-violating TPs in $B \to K^* (\to K \pi) \mu^+ \mu^-$ and $B \to
D^* (\to D \pi) \tau^- (\to \pi^- \nu_\tau) {\bar\nu}$ may be useful.

Of course, Belle II and LHCb will continue to measure CP violation in
$B$ decays. There may yet be surprises.

\begin{acknowledgments}
I congratulate the FPCP2020 organizers for their ability to adapt to
extremely trying circumstances. The transition from a regular ``live''
conference to one that was entirely on-line must have been exceedingly
difficult, but you made it appear seamless. Thank you. This work was
financially supported in part by NSERC of Canada.
\end{acknowledgments}

\bigskip 

\end{document}